# Fast 3D Variable-FOV Reconstruction for Parallel Imaging with Localized Sensitivities

Yigit Baran Can[1,2], Efe Ilicak[1,2], Tolga Çukur[1,2,3] PhD

[1]Department of Electrical and Electronics Engineering, Bilkent University, Ankara, TR-06800, Turkey
[2]National Magnetic Resonance Research Center, Bilkent University, Ankara, TR-06800, Turkey
[3]Neuroscience Program, Bilkent University, Ankara, TR-06800, Turkey

**Synopsis:** Several successful iterative approaches have recently been proposed for parallel-imaging reconstructions of variable-density (VD) acquisitions, but they often induce substantial computational burden for non-Cartesian data. Here we propose a generalized variable-FOV PILS reconstruction 3D VD Cartesian and non-Cartesian data. The proposed method separates k-space into non-intersecting annuli based on sampling density, and sets the 3D reconstruction FOV for each annulus based on the respective sampling density. The variable-FOV method is compared against conventional gridding, PILS, and ESPIRiT reconstructions. Results indicate that the proposed method yields better artifact suppression compared to gridding and PILS, and improves noise conditioning relative to ESPIRiT, enabling fast and high-quality reconstructions of 3D datasets.

**Introduction:** Several successful approaches based on image-domain (SENSE) or k-space (e.g., GRAPPA) operations have been previously proposed for parallel imaging reconstructions of variable-density acquisitions[1-3]. However, these methods involve iterative procedures for non-Cartesian data, resulting in substantial computational burden[4-6] in particular for three-dimensional (3D) reconstructions. An efficient k-space method based on PILS was recently proposed for fast reconstructions of 2D non-Cartesian data[7]. Here we propose a generalized, more efficient image-domain implementation for 3D variable-density Cartesian and non-Cartesian data. The proposed image-domain variable-FOV method is compared against conventional gridding, PILS, and ESPIRiT reconstructions. Simulations and in vivo results clearly show that the proposed method provides fast, high-SNR, artifact-suppressed reconstructions.

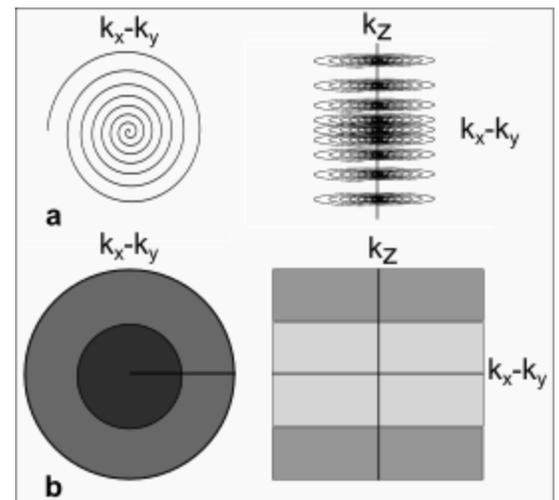

**Figure 1. a)** kx-ky plane projection and 3D appearance of 3D VD stack-of-spirals and **b)** separation of the trajectory into 4 non-intersecting annuli. The $k_x$-$k_y$ plane is divided into 2 annuli (inner and outer), and the $k_z$ direction is also divided into 2 annuli (central versus peripheral $k_z$ regions).

**Methods:** The proposed variable-FOV (var-FOV) method comprises three stages of processing for non-Cartesian data: individual-coil gridding reconstructions, frequency-dependent FOV restriction, and sensitivity-weighted coil combination. First, data from each coil are gridding reconstructed within the full FOV. Afterwards, the gridded image for each coil is filtered into N separate images, each corresponding to a different band of spatial frequencies in k-space. For example, for the 3D variable-density stack-of-spirals sampling trajectory in Figure 1, the kx-ky plane is divided into 2 annuli (inner and outer), and the kz direction is also divided into 2 annuli (central versus peripheral kz regions). When these divisions are considered together, they yield 4 non-overlapping bands of spatial frequency in k-space. The reconstruction FOV (FOV$_{recon}$) for each band is then restricted in accordance with the average sampling density within the band. In the final stage, coil sensitivities are estimated from the densely-sampled central part of k-space[8]. Based on these sensitivity estimates, coil-combination weights are then calculated separately for each frequency band, considering the varying FOV$_{recon}$. The individual-coil images are then linearly combined with these weights. The reconstructed image is: $P = \sum_{b=1}^{F}\sum_{i=1}^{C} K_{bi}W_{bi}$, where $K_{bi} = S_{bi}M_{bi}$ and $W_{bi} = M_{bi}D_i^*/\sqrt{(\sum_{i=1}^{C} M_{bi}|D_i|^2)}$. For frequency band b and coil i, $S_{bi}$ denotes the initial image obtained via gridding-reconstruction, $M_{bi}$ denotes the windowing function that restricts $FOV_{recon}$, and $W_{bi}$ denotes the coil combination weight. $D_i$ represents the coil-sensitivity estimate for coil i obtained from central k-space data.

The var-FOV method was demonstrated on 3D GRE acquisition of the knee[9] (0.5 mm isotropic) and a T1-weighted acquisition of the brain[10-14] (1mm isotropic). The datasets were resampled on 3D stack-of-spirals trajectories of varying acceleration factors (R). $FOV_{acq}$ was designed to linearly decrease from 10 cm to (5, 3, 1) cm for the knee and 18 to (9, 4, 2) cm for the brain. The trajectories had R = (2.6, 3.2, 4.1) and (10,8,6) interleaves for the knee, and R = (2.5, 3.7, 4.2) and (12, 8, 7) interleaves for the brain. The reconstructed images have matrix size of (320x320x220) and (240x240x240), for knee and brain respectively. SOS reconstructions were obtained via sum-of-squares combination of gridding reconstructions within the full FOV. PILS reconstructions were obtained via a sum-of-squares combination after individual-coil reconstructions were restricted to $FOV_{recon}$ supported by the highest sampling density in the trajectory. ESPIRiT reconstructions were obtained using a kernel size of [6, 6, 6], a Tykhonov parameter of 0.01 for sensitivity estimation, and an eigenvalue threshold of 0.001.

**Results:** Fig.2 shows the reconstructions at R=4.1 and 4.2 for the knee and brain datasets, respectively. For both datasets, SOS reconstructions suffer from significant residual aliasing artifacts and noise. While PILS offers somewhat improved artifact suppression, residual swirling artifacts can still be observed in the reconstructions because $FOV_{recon} > FOV_{acq}$ at high spatial frequencies. As expected, ESPIRiT is more effective than SOS and PILS in suppression of aliasing artifacts. However, ESPIRiT has relatively lower image SNR particularly at high R since it aims to solve a linear inverse problem. In contrast, the proposed var-FOV method attains a similar level of artifact suppression to ESPIRiT, while offering visibly improved SNR levels. This is because low frequency data is averaged over a larger extent and high frequency is restricted to regions with high sensitivity. These observations are supported by PSNR measurements for the knee and brain listed in Table I and Table II. For all R, Var-FOV outperforms SOS and PILS methods, while maintaining similar values compared to ESPIRiT. Var-FOV (Matlab implementation) reconstructed the knee and brain datasets in 293s and 190s, while ESPIRiT (C implementation) reconstructions took 730s and 382s, respectively.

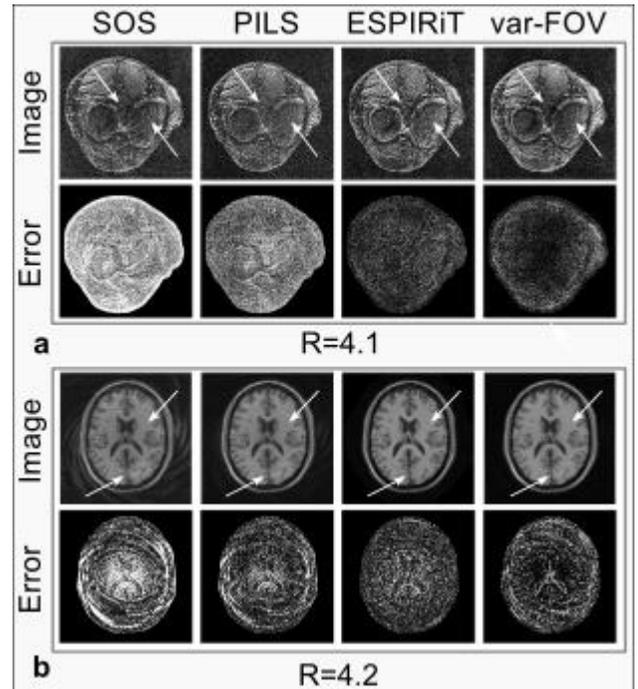

**Figure 2.** SOS, PILS, ESPIRiT and var-FOV reconstructions for **a)** knee and **b)** brain datasets. Arrows point at the aliasing artifacts. Error maps are scaled to [0-0.01]. SOS reconstructions suffer from significant aliasing artifacts and noise. While PILS yields relatively improved artifact suppression, residual swirling-like artifacts are still observed. ESPIRiT is more effective than both SOS and PILS in artifact suppression. However, ESPIRiT has relatively lower image SNR. In contrast, the var-FOV method attains a similar level of artifact suppression to ESPIRiT, while offering visibly improved SNR levels.

**Conclusion:** Providing high SNR and artifact-suppressed reconstructions of 3D variable-density non-Cartesian data, the proposed variable-FOV method is a fast and computationally efficient alternative to existing iterative approaches. It has fewer parameters that require tuning compared to regularized reconstructions. As a result, the proposed method enables a simple and efficient trade-off between artifact suppression and SNR.

**Table I. PSNR measurements on brain data**

|       | SOS          | PILS         | ESPIRiT      | Var-FOV      |
|-------|--------------|--------------|--------------|--------------|
| R=2.5 | 22.1 ± 0.143 | 25.1 ± 0.195 | 25.9 ± 0.236 | 27.0 ± 0.115 |
| R=3.7 | 21.6 ± 0.178 | 24.6 ± 0.240 | 25.7 ± 0.157 | 26.6 ± 0.100 |
| R=4.2 | 21.0 ± 0.220 | 24.4 ± 0.177 | 25.5 ± 0.275 | 26.2 ± 0.088 |

PSNR and SSIM measurements on SOS, PILS, ESPIRIT and var-FOV reconstructions of brain data. PSNR in dB is listed as mean ± s.e. across 16 axial slices spanning the volume.

**Table II. PSNR measurements on knee data**

|       | SOS          | PILS         | ESPIRiT      | Var-FOV      |
|-------|--------------|--------------|--------------|--------------|
| R=2.6 | 20.3 ± 0.094 | 22.6 ± 0.085 | 27.4 ± 0.104 | 28.0 ± 0.145 |
| R=3.2 | 21.2 ± 0.081 | 23.6 ± 0.067 | 27.0 ± 0.124 | 27.7 ± 0.216 |
| R=4.1 | 20.1 ± 0.054 | 22.4 ± 0.045 | 26.2 ± 0.159 | 26.4 ± 0.233 |

PSNR and SSIM measurements on SOS, PILS, ESPIRIT and var-FOV reconstructions of knee data. PSNR in dB is listed as mean ± s.e. across 16 axial slices spanning the volume.

**References**

[1] Pruessmann KP, Weiger M, Scheidegger MB, Boesiger P. SENSE: Sensitivity encoding for fast MRI.Magn Reson Med. 1999; 42(5): 952–962.

[2] Griswold MA, Jakob PM, Heidemann RM, Nittka M, Jellus V, Wang J, Kiefer B, Haase A. Generalized autocalibrating partially parallel acquisitions (GRAPPA) Magn Reson Med. 2002; 47(6): 1202–1210.

[3] Uecker M, Lai P, Murphy MJ, Virtue P, Elad M, Pauly JM, Vasanawala SS, Lustig M. ESPIRiT--an eigenvalue approach to autocalibrating parallel MRI: where SENSE meets GRAPPA. Magn Reson Med. 2014; 71(3): 990-1001.

[4] Pruessmann KP, Weiger M, Börnert P, Boesiger P. Advances in sensitivity encoding with arbitrary k-space trajectories. Magn Reson Med. 2001; 46(4): 638–51.

[5] Griswold M, Heidemann R, Jakob P. Direct parallel imaging reconstruction of radially sampled data using GRAPPA with relative shifts; Proc of the 11th Annual Meeting of the Int Soc Magn Reson Med; Toronto, Canada. 2003; p. 2349.

[6] Heidemann RM, Griswold MA, Seiberlich N, Krüger G, Kannengiesser SAR, Kiefer B, Wiggins G, Wald LL, Jakob PM. Direct parallel image reconstructions for spiral trajectories using GRAPPA. Magn Reson Med. 2006; 56(2): 317–26.

[7] Cukur T, Santos JM, Pauly J, Nishimura D Variable-Density Parallel Image Acquisition and Reconstruction with Partially Localized Coil Sensitivities; Proc of the 15th Annual Meeting of the Int Soc Magn Reson Med; Berlin, Germany. 2007; p. 178.

[8] Bydder M, Larkman DJ, Hajnal JV. Combination of signals from array coils using image-based estimation of coil sensitivity profiles. Magn Reson Med. 2002; 47: 539-548.

[9] ISMRM, "MRI Unbound Simulated Data," 2010. [Online]. Available: http://www.ismrm.org/mri_unbound/simulated.htm. [Accessed 26 11 2016].

[10] McGill, "BrainWeb: Simulated MRI Volumes for Normal Brain," [Online]. Available: http://brainweb.bic.mni.mcgill.ca/brainweb/selection_normal.html. [Accessed 26 11 2016].

[11] C.A. Cocosco, V. Kollokian, R.K.-S. Kwan, A.C. Evans : "BrainWeb: Online Interface to a 3D MRI Simulated Brain Database" NeuroImage, vol.5, no.4, part 2/4, S425, 1997 -- Proceedings of 3-rd International Conference on Functional Mapping of the Human Brain, Copenhagen, May 1997.

[12] R.K.-S. Kwan, A.C. Evans, G.B. Pike : "MRI simulation-based evaluation of image-processing and classification methods" IEEE Transactions on Medical Imaging. 18(11):1085-97, Nov 1999.

[13]  R.K.-S. Kwan, A.C. Evans, G.B. Pike : "An Extensible MRI Simulator for Post-Processing Evaluation" Visualization in Biomedical Computing (VBC'96). Lecture Notes in Computer Science, vol. 1131. Springer-Verlag, 1996. 135-140.

[14]  D.L. Collins, A.P. Zijdenbos, V. Kollokian, J.G. Sled, N.J. Kabani, C.J. Holmes, A.C. Evans : "Design and Construction of a Realistic Digital Brain Phantom" IEEE Transactions on Medical Imaging, vol.17, No.3, p.463--468, June 1998.